# Optically reconfigurable quantum spin-valley Hall effect of light in coupled nonlinear ring resonator lattice

Haofan Yang[1], Jing Xu[1, *], Zhongfei Xiong[1], Xinda Lu[1], Ruo-Yang Zhang[2], Yuntian Chen[1, †], and Shuang Zhang[3, 4‡]

[1]*School of Optical and Electronic Information & Wuhan National Laboratory for Optoelectronics, Huazhong University of Science and Technology, Wuhan 430074, China*

[2]*Department of Physics, The Hong Kong University of Science and Technology, Clear Water Bay, Hong Kong*

[3]*Department of Physics, University of Hong Kong, Hong Kong, China*

[4]*Department of Electrical & Electronic Engineering, University of Hong Kong, Hong Kong, China*

* Corresponding author: jing_xu@hust.edu.cn
† Corresponding author: yuntian@hust.edu.cn
‡ Corresponding author: shuzhang@hku.hk

*Abstract*. Scattering immune propagation of light in topological photonic systems may revolutionize the design of integrated photonic circuits for information processing and communications. In optics, various photonic topological circuits have been developed, which were based on classical emulation of either quantum spin Hall effect or quantum valley Hall effect. On the other hand, the combination of both the valley and spin degrees of freedom can lead to a new kind of topological transport phenomenon, dubbed quantum spin valley Hall effect (QSVH), which can further expand the number of topologically protected edge channels and would be useful for information multiplexing. However, it is challenging to realize QSVH in most known material platforms, due to the requirement of breaking both the (pseudo-)fermionic time-reversal ($\mathcal{T}$) and parity symmetries ($\mathcal{P}$) individually, but leaving the combined symmetry $\mathcal{S} \equiv \mathcal{TP}$ intact. Here, we propose an experimentally feasible platform to realize QSVH for light, based on coupled ring resonators mediated by optical Kerr nonlinearity. Thanks to the inherent flexibility of cross-mode modulation (XMM), the coupling between the probe light can be engineered in a controllable way such that spin-dependent staggered sublattice potential emerges in the effective Hamiltonian. With delicate yet experimentally feasible pump conditions, we show the existence of spin valley Hall induced topological edge states. We further demonstrate that both degrees of freedom, i.e., spin and valley, can be manipulated simultaneously in a reconfigurable manner to realize spin-valley photonics, doubling the degrees of freedom for enhancing the information capacity in optical communication systems.

*Introduction.* The interplay among the various degrees of freedom of photons, i.e., photon spin, valley, and sublattice pseudospin in a planar honeycomb structures provide a rich playground for realizing different Hall effects of light, including photonic spin Hall effect [1] and photonic valley Hall effect [2], that feature topologically protected edge states. Those intriguing phenomena essentially rely on carefully engineered Berry curvature distributed over different valleys and spin sectors [2-20]. For instance, in the presence of time-reversal symmetry, one can realize either quantum spin Hall effect [6,14] via spin-orbit interaction, or quantum valley Hall effect [2,3,21] via breaking inversion symmetry with external biased field. Interestingly, the combination of both the valley and spin degrees of freedom, i.e., quantum spin valley Hall effect (QSVH) [22,23], has been identified to have a deep connection to the antiferromagnetic ordering [24] in the 2D materials, which may lead to far-reaching implications and applications in spintronics. Moreover, QSVH of light can potentially double the number of topologically protected channels for increasing the information capacity in optical communication and bit data transmission. However, it is very challenging to realize QSVH due to the requirement of violating both (pseudo-)fermionic time-reversal symmetry ($\mathcal{T}$) and parity ($\mathcal{P}$) individually, but preserving the symmetry of the joint operation $\mathcal{S} \equiv \mathcal{TP}$ [24,25]. There are a few attempts to realize QSVH in condensate matter physics, including the usage of monolayer antiferromagnetic material [24] or single-layer graphene with an in-plane applied magnetic field [26], or AA-stacked bilayer graphene [27]. Notably, Gladstone et al. recently proposed a photonic graphene based on electromagnetic metamaterial to realize the spin-valley photonic topological insulators via accidental degeneracy. However, the required fine tuning of structural parameters hinders the experimental realization of QSVH at optical frequencies [28].

In this work, we propose an experimentally feasible setup to realize optically reconfigurable QSVH at optical frequencies, by exploiting the Kerr nonlinear effect in coupled ring resonators on a honeycomb lattice. Note that optical Kerr nonlinearity has been utilized extensively in the context of topologic photonics [29-36]. Our investigation provides a route towards the integration of the valley and spin degrees of freedom for information transport and processing in integrated photonic circuits, as well as the possibility of exploring antiferromagnetic ordering in planar photonic platform. An exemplary system for realizing QSVH is shown in Fig. 1(a). The system consists of a 2D honeycomb array of ring resonators. Due to the rotational symmetry, each ring resonator supports a pair of circulating modes, i.e., clockwise (*CW*) and counterclockwise (*CCW*) modes. To realize QSVH for the probe light, the system is designed in such a way that each circulating mode in ring *a* can only couple to the same (reverse) mode in ring *b* for the pump (probe) light. This coupling scheme can be achieved for carefully chosen wavelengths of the pump and probe light by introducing an intermediate resonator between two adjacent rings, i.e., the small gray ring between ring *a* and ring *b*, as shown by the inset of Fig. 1(a) [see S3 in supplementary materials (SM) for details].

*Coupled Nonlinear Ring Resonator Lattice.* We begin with the analysis of the cross-mode modulation (XMM) [37,38] between two pump lights (*CW* with complex amplitude of $p$ and *CCW* with complex amplitude $q$ at the same frequency $f_1$) and two probe lights (*CW* and *CCW* modes at frequency $f_2$) within a single isolated ring, i.e., the target ring. The XMM occurs in any materials with $\chi^{(3)}$ nonlinearity since the phase-matching condition is automatically satisfied [37,38]. Under the initial pump condition $\left|\varphi_{pump}^0\right\rangle = \sqrt{P_p}(p \quad q)$ with a total pump power $P_p$, *CW* and *CCW* modes of the pump light remain decoupled in the presence of optical nonlinearity (see S1 in SM for details). Assuming that the probe beam is much weaker than the pump beam, the Hamiltonian of the probe light is completely determined by the pump light [37-39], which reads,

$$H_p = 2\gamma P_p v_p \begin{pmatrix} 1 & pq^* e^{i\Delta\zeta t} \\ p^* q e^{-i\Delta\zeta t} & 1 \end{pmatrix} \tag{1}$$

where $\gamma$ is the nonlinear parameter, $v_p$ is the phase velocity of pump light, $p$ and $q$ are the normalized complex amplitudes of *CW* and *CCW* ($|p|^2 + |q|^2 = 1$) for the pump light at $t = 0$, respectively, and $\Delta\zeta$ is the frequency offset between *CW* and *CCW* pump light in presence of XMM.

The *CW* or *CCW* mode of the pump light inside a ring resonator (any ring at the outermost of the lattice) can be excited by a nearby waveguide, as shown in Fig. 1(a), depending on whether the pumping beam is launched from the left- or right- hand side of the waveguide. The relative complex amplitude of *CW* mode and *CCW* mode of the pump light are therefore determined by the amplitudes and phases of the waveguide inputs from both ends. Once the target ring lights up under the pump condition $\sqrt{P_p}(p \quad q)$, all other rings on the lattice are excited by evanescent coupling via the intermediate resonators with the same amplitude apart from additional propagation phase $\phi$, i.e., $\sqrt{P_p}(p^{i\phi} \quad q^{i\phi})$ due to the fact that the gray intermediate ring only couples the modes of the same helicity between adjacent lattice rings *a* and *b*, see S2 and S3 in SM for detailed analysis. Consequently, the tight binding Hamiltonian of the probe light is given as follows,

$$\begin{aligned} H = &\sum_i (V_1 a_{i,1}^+ a_{i,1} + V_2 a_{i,2}^+ a_{i,2}) + \sum_i (V_1 b_{i,1}^+ b_{i,1} + V_2 b_{i,2}^+ b_{i,2}) \\ &+ \sum_{\langle ij \rangle} \chi \left\{ (a_{i,1}^+ b_{j,2} + c.c.) + (a_{i,2}^+ b_{j,1} + c.c.) \right\} \\ &+ \sum_i (g a_{i,1}^+ a_{i,2} + c.c.) + \sum_i (g b_{i,1}^+ b_{i,2} + c.c.) \end{aligned} \quad (2)$$

where $a_{i,1/2}/a_{i,1/2}^+$ and $b_{i,1/2}/b_{i,1/2}^+$ denote the creation/annihilate operators for *CW/CCW* mode in ring *a* and ring *b* on lattice site *i*, respectively, *g* represents the nonlinear coupling induced by pump light which is given by $g = 2\gamma v_p P_p p q^* e^{i\Delta\zeta t}$, $V_1 = V_2 = 2\pi f_2 + 2\gamma v_p P_p$, and $\chi$ the linear coupling strength. The third term in *H* sums over the nearest-neighbor sites.

With substitution $m = 2\gamma v_p P_p p q^*$ and $\omega = \Delta\zeta$, under the basis of $\psi_k = (a_1 \ b_2 \ a_2 \ b_1)^T$, the Hamiltonian *H* can be reformulated in k-space as [40],

$$\mathcal{H}(\vec{k}) = \begin{pmatrix} V_1 & \alpha & me^{i\omega t} & 0 \\ \alpha^* & V_1 & 0 & m^* e^{-i\omega t} \\ m^* e^{-i\omega t} & 0 & V_1 & \alpha \\ 0 & me^{i\omega t} & \alpha^* & V_1 \end{pmatrix}, \quad (3)$$

where $\alpha = \chi(e^{-i\vec{k}\cdot\vec{e}_1} + e^{-i\vec{k}\cdot\vec{e}_2} + e^{-i\vec{k}\cdot\vec{e}_3})$ with $\vec{e}_i (i=1,2,3)$ being the reciprocal lattice vectors connecting a ring of *a* to its three nearest neighboring rings of *b*, respectively. The Hamiltonian $\mathcal{H}(\vec{k})$ is periodic in time, and satisfies $\mathcal{H}(t+T) = \mathcal{H}(t)$, leading to the time-dependent Schrödinger equation $i\partial_t |\psi(t)\rangle = \mathcal{H}(t)|\psi(t)\rangle$ [41,42]. By following the standard procedure of solving the Floquet problem [43], we obtain $\sum_N (\mathcal{H}_{M-N} - M\omega\delta_{MN})|\Phi^N\rangle = \varepsilon|\Phi^M\rangle$, where $\mathcal{H}_M = \frac{1}{T}\int_0^T \mathcal{H}(t) e^{iM\omega t} dt$ is the *M*-th Fourier harmonic of

$\mathcal{H}(\vec{k})$. Given a large $\omega$ and small $\mathcal{H}_{+1,-1}$ [43-45], the effective Hamiltonian is well-approximated by $\mathcal{H}_{eff} = \mathcal{H}_0 - \frac{1}{\omega}[\mathcal{H}_{+1}, \mathcal{H}_{-1}] \approx V_1 I + \mathcal{H}_e$, where $\mathcal{H}_e$ is

$$\mathcal{H}_e = \begin{pmatrix} \frac{1}{\omega}|m|^2 & \alpha & 0 & 0 \\ \alpha^* & -\frac{1}{\omega}|m|^2 & 0 & 0 \\ 0 & 0 & -\frac{1}{\omega}|m|^2 & \alpha \\ 0 & 0 & \alpha^* & \frac{1}{\omega}|m|^2 \end{pmatrix}. \quad (4)$$

The effective Hamiltonian does not change if additional high-frequency expansion terms are included, see S5 in SM for details. Interestingly, Eq. (4) is reminiscent of the antiferromagnetic-like spin-staggered sublattice potential emerged in the lattice Hamiltonian, which violates both (pseudo-)fermionic time-reversal symmetry ($\mathcal{T}$) and parity ($\mathcal{P}$) individually but preserves the symmetry of the joint operation $\mathcal{S} \equiv \mathcal{TP}$ [24]. In the following, we will discuss the underlying mechanisms for achieving the spin-valley Hall effect.

*Topological invariances.* As illustrated in Fig. 1(a), the static coupling for the probe light only occurs between the modes of opposite circulation at the nearest neighbor ring resonators. Thus, it is convenient to define CW(*a*)+CCW(*b*) and CCW(*a*)+CW(*b*) as spin up $\psi_\uparrow$ and spin down $\psi_\downarrow$, respectively (see Fig. 1(b)). Without pumping, the band structure of the probe light is shown in Fig. 1(c), which exhibits Dirac cones at the points K and K′, corresponding to $K_\pm = (\pm\frac{4}{3\sqrt{3}}\frac{\pi}{d}, 0)$ in the reciprocal space, where *d* is the length of the nearest-neighbor bonds. In the presence of pumping with $|\varphi_{pump}^0\rangle = \sqrt{P_p}(p \quad q)$, which is coined type-*A* configuration here, the nonlinear interaction introduces opposite mass terms to the two block-diagonalized spin sectors. As such, the degeneracies at the valleys K and K′ are lifted with a gap $\Delta = \frac{2|m|^2}{\omega}$, where $m = m_0 v_p$ and $\omega = 5 m_0 v_p$ with $m_0 = 0.04 \text{m}^{-1}$ (see the details in SM S3). One notes that the two spin sectors can be used to realize the spin valley Hall effect if another pumping configuration (termed type-*B*) is introduced by simply interchanging the pump conditions of *CW* and *CCW* modes in

type-$A$ configuration, i.e., $\left|\varphi_{pump}^{0}\right\rangle = \sqrt{P_p}(q\ \ p)$, which leads to interchanged mass terms between the two spin states. It is noted that type-$A$ and type-$B$ configurations share the same bulk band structure, as shown in Fig. 1(d). However, the band topology is different, where the topological transition occurs as the pump condition is switched from type-$A$ structure to type-$B$ configuration. This can be seen from the effective Hamiltonian near $K_\pm$ for type-$A$, e.g., for the spin-up state $\psi_\uparrow$, $H_{K/K'}^{\psi_\uparrow}(l) = \frac{3}{2}\chi d\left(\pm l_x \sigma_x + l_y \sigma_y\right) + n\sigma_z$ ($\sigma_i (i=x,y,z)$ are the Pauli matrices), in which $n = \pm\frac{|m|^2}{\omega}$ being positive for type $A$ and negative for type $B$. The local Berry curvature ($z$ component) of the lower band for spin up states at the $K/K'$ valley can be approximated as $\Omega_{K/K'}^{\psi_\uparrow} = \pm\frac{n}{2(\delta k^2 + n^2)^{3/2}}$ (the offset $\delta k = \frac{3}{2}\chi d|l|$) [8], as shown in Fig. 1(e) for both type-$A$ and type-$B$ configurations. Accordingly, we can calculate the valley Chern numbers of the first band, i.e., for spin up states $\psi_\uparrow$ at a single valley, by integrating the local Berry curvature derived from the effectively Hamiltonian [8,23]

$$C_{K/K'}^{\psi_\uparrow} = \frac{1}{2\pi}\int_{HBZ} \Omega_{K/K'}^{\psi_\uparrow}(\delta k) dS = \pm\frac{1}{2}\text{sgn}(n) \tag{5}$$

where $HBZ$ denotes the half of the Brillouin zone containing the $K/K'$ valley. Eq. (5) shows that for each valley, the valley Chern number is $\pm\frac{1}{2}$ with the sign solely determined by the sign of $n$. The calculated valley Chern numbers in the $\psi_\uparrow$ sector for type $A$ and type $B$ configurations are summarized in Fig. 1(e), while a similar procedure applies to the spin-down states $\psi_\downarrow$. Although the total Chern number in each spin sector or each valley sector is trivial, a novel topological invariant known as spin-valley Chern number $C_{sv} = \frac{1}{2}\left(C_K^{\psi_\uparrow} - C_{K'}^{\psi_\uparrow} - C_K^{\psi_\downarrow} + C_{K'}^{\psi_\downarrow}\right)$ [23] can be introduced to characterize the nontrivial band topology of the type $A$ and type $B$ structures, i.e., $C_{sv}^A = 1$ for type $A$ and $C_{sv}^B = -1$ for type $B$. The non-zero spin-valley Chern number indicates that a special kind of spin-valley polarized interface transportation, i.e., spin valley Hall effect, which is distinct from spin Hall effect and valley Hall effect, can be achieved by stacking $A$ and $B$ structures together.

***Spin- and valley- polarized edge states.*** For a domain wall between type-*A* and type-*B* configurations, the change of the spin-valley Chern numbers across the interface for the first band is given as $|\Delta C_{sv}| = |C_{sv}^{A(B)} - C_{sv}^{B(A)}| = 2$. Dictated by the bulk-edge correspondence, non-trivial edge states emerge at the domain wall [4,5]. Here the domain wall is formed by removing a line of intermediate rings at the interface of type *A* (upper) and type *B* (bottom) domains as *AB*-type, or other way around as *BA*-type, as shown by the inset of Fig. 2(a). The band structures under $m = m_0 v_p$ and $\omega = 5 m_0 v_p$ for both spin sectors are shown in Figs. 2(a) and 2(b) respectively, with the edge states in *AB*-type (*BA*-type) domain walls represented by the blue (red) lines. For each spin sector, valley-polarized topological edge states emerge at the domain wall, while the chiralities of the edge states of different spins are opposite, thus confirming the presence of the spin-valley Hall effect of light, as illustrated by the spin up state $\psi_\uparrow$ in Figs. 2(c) and 2(d) and the spin down state $\psi_\downarrow$ in Figs. 2(e) and 2(f), where the excitation of the probe light can be experimentally implemented by introducing a circular grating coupler [46], see S4 in SM. For $\psi_\downarrow$ ($\psi_\uparrow$) polarizations at the K valley, the edge state of the AB-type domain wall is backward (forward) propagating, while the propagation direction is reversed at the K' valley, which is consistent with the band structure in Figs. 2(a) and 2(b). Moreover, at each valley, i.e., Figs. 2(c) and 2(e) for K (Figs. 2(d) and 2(f) for K'), the two spin-polarized topological edge states propagate along opposite directions due to the spin-momentum locking, as shown in Fig. 2(e). This is a manifestation of both the spin- and valley-chirality in the spin-valley Hall effect [22,23], with the total number of the edge states agreeing well with the valley Chern number for both spins, $\Delta C_{AB,K}^{\psi_\downarrow} = C_{A,K}^{\psi_\downarrow} - C_{B,K}^{\psi_\downarrow} = -1$ ($\Delta C_{AB,K}^{\psi_\uparrow} = 1$) and $\Delta C_{AB,K'}^{\psi_\downarrow} = C_{A,K'}^{\psi_\downarrow} - C_{B,K'}^{\psi_\downarrow} = 1$ ($\Delta C_{AB,K'}^{\psi_\uparrow} = -1$). Conveniently, the *BA*-type domain wall can be realized by flipping the initial input pump condition of the input waveguide in the *AB*-type, as shown in Fig. 1(a). In the *BA*-type structure, the propagating directions of the edge state for each valley and spin are exactly opposite to those of *AB*-type. Thereby, the presence of spin- and valley- polarized edge states at the *AB*- and *BA*- type domain walls unambiguously shows the existence of spin-valley Hall effect of light, to the best of our knowledge, which has never been reported in photonic platforms.

To understand the back scattering properties and intervalley scattering behavior of the spin-valley-polarized edge states, we examine the edge state propagation along a zigzag domain wall, as illustrated by Fig. 3(a). A point source (the red pentagram) is placed at the left end of the domain wall to excite the valley-polarized edge states. Fig. 3(a) shows that the topological valley edge states of the $\psi_\uparrow$ and $\psi_\downarrow$ can be guided around the zigzag path smoothly. As mentioned above, there are two robust edge states propagating along the same direction along the domain wall, corresponding to two different spin/valley combinations. Since the two combinations in any ring correspond to the two modes with opposite circulations, the spin-orbit coupling of the waveguide modes and the ring resonances can be used to spatially separate the two spins, i.e., $\psi_\uparrow$ (the white arrow) and $\psi_\downarrow$ (the yellow arrow), separately, as shown in Fig. 3(b). Therefore, both degrees of freedom, i.e., spin and valley, can be manipulated simultaneously by introducing two waveguides on the left and right ports of the domain wall, see S6 in SM.

*Discussion and conclusion.* In summary, we have proposed an optically reconfigurable spin-valley Hall effect based on Kerr nonlinearity in an experimentally feasible integrated platform. On a photonic honeycomb lattice formed by coupled ring resonators, we show that topologically protected spin-valley-polarized edge states can be selectively excited, and spatially separated in terms of the spin and valley indices. In contrast to spin Hall effect or valley Hall effect, our work takes advantage of the spin and valley degrees of freedom simultaneously that double the number of topologically protected information channels for future integrated photonics. Besides, our research suggests that the optical nonlinearity, especially XMM can be used as flexible means to optically tune the Hamiltonian of the probe light, as illustrated by the dynamic switching between *AB*- and *BA*- type domain walls simply by changing the input initial pump condition of the same structure. Importantly, by the nature of XMM, such type of nonlinear effect occurs in any $\chi^{(3)}$ nonlinear optical material due to the automatic phase-matching, which might be useful in the realization of the reconfigurable topological photonics merely using optical means.

This work is supported by the National Key Research and Development Program of China (2019YFB2203103), National Natural Science Foundation of China (NSFC) (no. 61775063, no. 61735006, and no. 11874026), and Horizon 2020 Action, projects 648783 (TOPOLOGICAL) and 734578 (D-SPA).

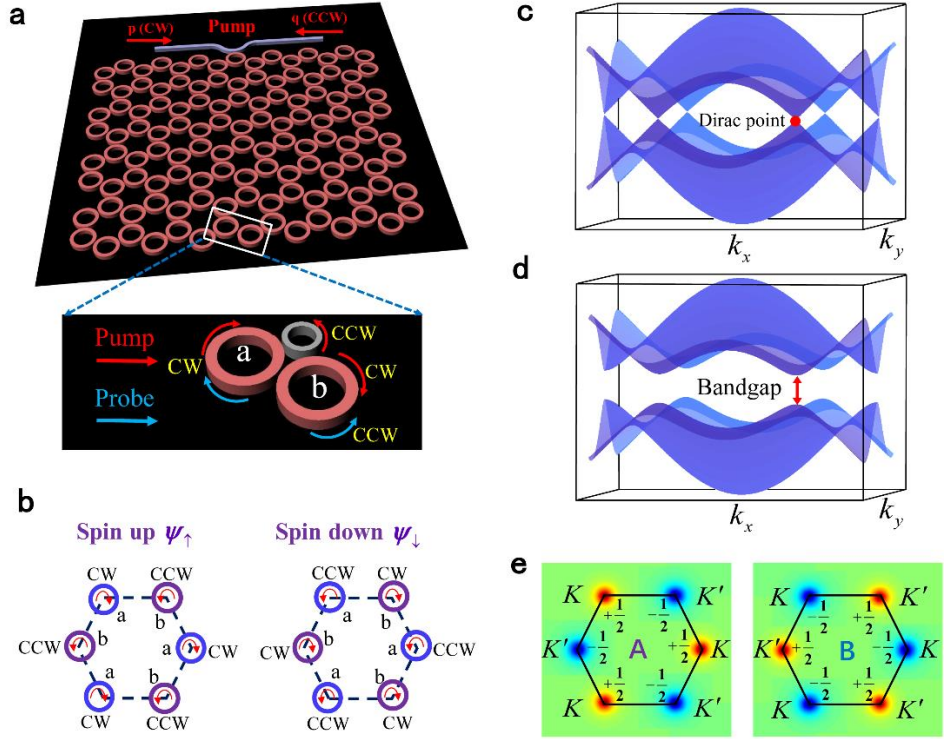

**FIG. 1.** A 2D honeycomb array of ring resonators and its bulk band structure. (a) A 2D honeycomb array of ring resonators. Each ring support supports *CW* and *CCW* mode. The illustration represents the pump-probe configuration and the corresponding mode coupling between the nearest neighbor ring. (b) The schematic diagram for spin up and spin down. (c) Band structure for the case of non-pump comprising a honeycomb lattice ( $P_p = 0$ ). Note the band crossings at the Dirac point (the red dot). (d) Band structure for the spin-valley Hall photonic topological insulator. Dirac points is to open a gap with magnitude $\Delta = \frac{2|m|^2}{\omega}$, which corresponds to the bandgap in a Floquet photonic topological insulator. (e) Calculated Berry curvature of the first band near the *K* and *K′* valleys and corresponding valley Chern numbers for type-*A* and type-*B* for the $\psi_\uparrow$.

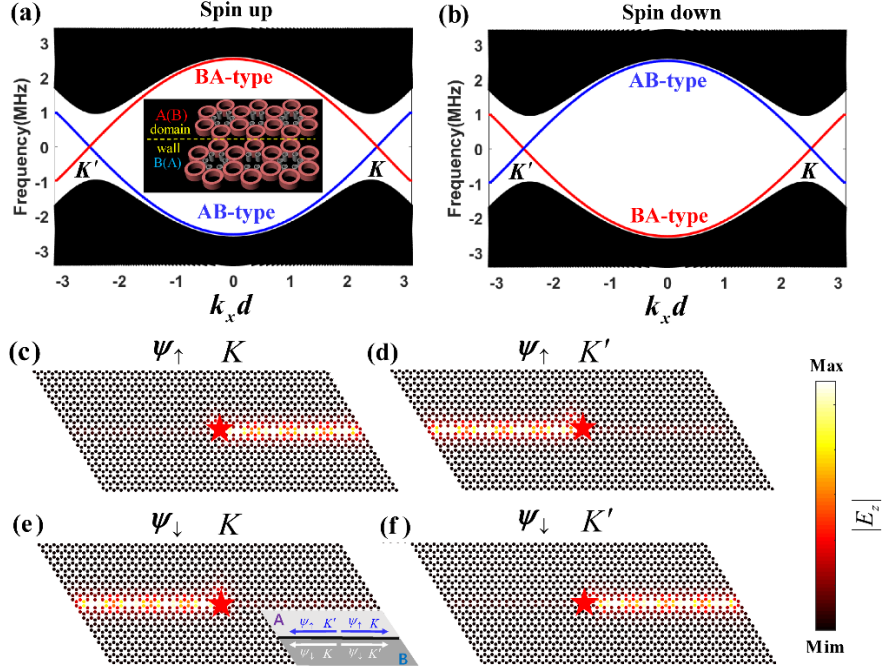

**FIG. 2.** Calculated band structure for spin up (a) and spin down (b) of an *AB*-type (blue lines) domain wall and a *BA*-type (red lines) domain wall. Frequency of the band structure corresponds to the frequency relative to that of the probe light. The curves correspond to the bulk states. The illustration in (a) is the schematic of a straight domain wall (yellow dashed line) between the upper domain with type *A*(*B*) and the lower domain with type *B*(*A*). One-way propagation valley edge states of *AB*-type domain wall (c-f). The propagating direction of excited edge states on (c) K valley of $\psi_\uparrow$; (d) K′ valley of $\psi_\uparrow$; (e) K valley of $\psi_\downarrow$ and (f) K′ valley of $\psi_\downarrow$ agree with prediction from the band structure. The illustration in (e) shows the schematic diagrams of the domain wall and corresponding edge states. The red pentagram represents the source whose phase matches with the edge states at the K/ K′ valley. The color indicates the amplitude of $E_z$ profiles.

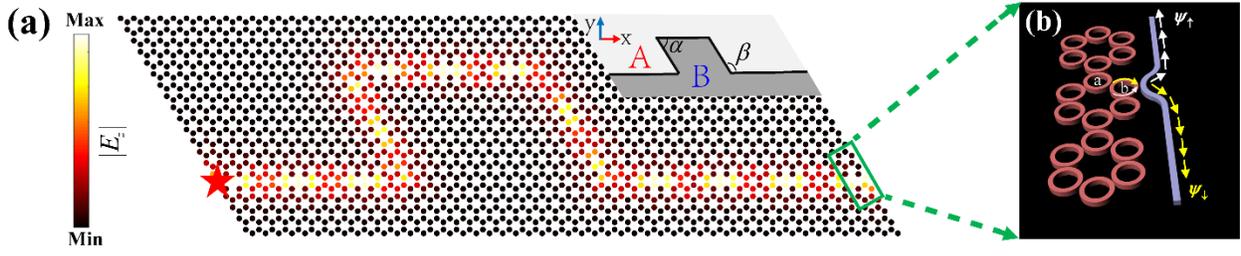

**FIG. 3.** Robustness of topological valley edge states of the $\psi_\uparrow$ and $\psi_\downarrow$ and the spatial separation of the two spins. (a) Simulated field map of $E_z$ profiles in the *xy* plane. The red pentagram indicates the location of the excitation source. The illustration represents the schematic of the zigzag domain wall (black solid line) between the upper domain with type *A* and the lower domain with type *B*. The corner angle $\alpha = 60°$ and $\omega = 120°$. The color indicates the amplitude of $E_z$ profiles. (b) Through the natural spin-orbit coupling of waveguide and ring, the two spins, i.e., $\psi_\uparrow$ (the white arrow) and $\psi_\downarrow$ (the yellow arrow) are separated.